\documentclass[twocolumn,english,aps,prb,groupedaddress]{revtex4-1}
\usepackage[T1]{fontenc}
\usepackage[latin9]{inputenc}
\usepackage{amstext}
\usepackage{graphicx}
\usepackage{esint}

\makeatletter

\providecommand{\tabularnewline}{\\}

 \@ifundefined{textcolor}{}
 {%
   \definecolor{BLACK}{gray}{0}
   \definecolor{WHITE}{gray}{1}
   \definecolor{RED}{rgb}{1,0,0}
   \definecolor{GREEN}{rgb}{0,1,0}
   \definecolor{BLUE}{rgb}{0,0,1}
   \definecolor{CYAN}{cmyk}{1,0,0,0}
   \definecolor{MAGENTA}{cmyk}{0,1,0,0}
   \definecolor{YELLOW}{cmyk}{0,0,1,0}
 }

\usepackage{multirow}%\usepackage{setspace} %remove later
\makeatother

\usepackage{babel}

\begin{document}

\title{Fermi level pinning can determine polarity in semiconductor nanorods}

\author{Philip W. \surname{Avraam}}

\author{Nicholas D. M. \surname{Hine}}

\author{Paul \surname{Tangney}}

\author{Peter D. \surname{Haynes}}

\email{p.haynes@imperial.ac.uk}

\affiliation{Department of Physics and Department of Materials, Imperial College
London, Exhibition Road, London SW7 2AZ, United Kingdom}

\date{\today}
\begin{abstract}
First-principles calculations of polar semiconductor nanorods reveal
that their dipole moments are strongly influenced by Fermi level pinning.
The Fermi level for an isolated nanorod is found to coincide with
a significant density of electronic surface states at the end surfaces,
which are either mid-gap states or band-edge states. These states
pin the Fermi level, and therefore fix the potential difference across
the rod. We provide evidence that this effect can have a determining
influence on the polarity of nanorods, with consequences for the way
a rod responds to changes in its surface chemistry, the scaling of
its dipole moment with its size, and the dependence of polarity on
its composition. 
\end{abstract}

\pacs{334}

\maketitle

\section{Introduction}

\label{sec:introduction}

Semiconductor nanostructures in solution are a very exciting class
of material due to our growing ability to manipulate their shapes
and sizes, and the superstructures into which they assemble, to produce
a wide range of technologically useful properties.\cite{smallisdifferent,X.Michalet01282005,NirTessler02222002,kazesetal,Wendy,Nieetal,shevchenko}

Nanocrystals of binary semiconductors, such as those of ZnO, have
been observed to exhibit very large dipole moments\cite{PhysRevLett.79.865,shim:6955,PhysRevLett.90.097402}
which affect their internal electronic structure (and therefore their
optical properties) as well as their interactions with their environment.
The latter may influence the kinetics of self-assembly and the stability
of the structures formed.\cite{talapin}

A detailed understanding of the factors contributing to this polarity
in nanocrystals has proven elusive\cite{Goniakowski} for two main
reasons: first, many factors are involved, ranging from surface chemistry,
to the non-centrosymmetric nature of the underlying crystal, to quantum
confinement, to long-range electrostatics, to interactions with the
solvent and considerations of thermodynamic stability; and second,
the limitations of current experimental techniques, which do not allow
the level of control over, or knowledge of, the state of the system,
that is necessary to be able to disaggregate these factors.

Computer simulation is an ideal tool for addressing this problem.\cite{rabani:5355,rabani:1493,Shanbhag:2006ix,wang}
Recent developments in linear-scaling density-functional theory (LS-DFT),
make accurate quantum-mechanical methods applicable to nanocrystals
of realistic sizes.

In our earlier work\cite{PhysRevB.83.241402} we presented results
from LS-DFT calculations using the \textsc{onetep} code,\cite{skylaris:084119,onetep-forces}
of the ground-state charge distributions in GaAs nanorods of sizes
comparable to those found in experiment. We found that its dipole
moment depends strongly on the surface termination, particularly of
its polar surfaces, with full hydrogen termination on polar surfaces
strongly reversing its direction.

A common feature of all of the nanorods studied was that the Fermi
energy was found to coincide with a significant density of states
located at the end surfaces of the rods. {\em Fermi level pinning}
(FLP) is known to occur in semi-infinite semiconductor surfaces when
states are found at the Fermi energy, and in this work we show that
a finite-surface version of FLP plays a crucial role in determining
the polar characteristics of such nanorods.

In section~\ref{sec:methods} we outline the simulation details and
methodology. In section~\ref{sec:FLP} we show that mid-gap states
on the end surfaces of the rod can pin the Fermi energy, which in
turn determines the potential difference across the nanorod, and therefore
its dipole moment.

In section~\ref{sec:ionic-charge} we take up an important observation
from our previous work, namely that nanorods terminated on their ends
with ions of very different ionic charge can nevertheless have very
similar dipole moments. This observation is particularly problematic
for simple ionic or bond-electron counting models,\cite{Goniakowski}
which can fail to predict the dipole moments as a result. These models
are not able to explain the magnitudes of the differences in polarity
between nanorods of different surface terminations. We show that our
FLP model can rationalize these observations.

In section~\ref{sec:NR-size} we calculate the variation of nanorod
polarization with rod length and cross-sectional area. The dipole
moment is found to increase with nanorod size in a manner consistent
with maintaining a `pinned' Fermi level at the end polar surfaces
of the nanorod.

Finally, in section~\ref{sec:NR-type} we study the variation in
polarity between nanorods of different compositions (specifically
GaAs, GaN and AlN), again illustrating the determining role of FLP
for the rod polarizations.

\section{\label{sec:methods} Simulation Methodology}

This work uses linear-scaling density-functional theory (LS-DFT) as
implemented in the \textsc{onetep} code.\cite{skylaris:084119,onetep-forces}
This method combines the benefits of linear scaling, in that computational
resources for calculating the total energy of an $N$-atom system
scales as $O(N)$, with the accuracy of plane-wave methods.\cite{onetep-pwaccuracy}
In \textsc{onetep} the single-particle density matrix is represented
by an optimized set of non-orthogonal, strictly localized, Wannier-like
orbitals $\{\phi_{\alpha}({\bf {r})\}}$, and is written \begin{equation}
\rho(\mathbf{r},\mathbf{r'})=\sum_{\alpha\beta}\phi_{\alpha}(\mathbf{r})K^{\alpha\beta}\phi_{\beta}^{*}(\mathbf{r'})\end{equation}

\noindent where $K^{\alpha\beta}$ is the \textit{density kernel}
representing a generalization of the occupation numbers to a non-orthogonal
basis. Both the local orbitals and the density kernel are optimized
during the calculation. The three tuneable parameters controlling
the quality of the representation are:\cite{0953-8984-17-37-012}
the `plane-wave' cutoff energy $E_{\text{cut}}$, defining the grid-spacing
for the grid on which the local-orbitals are represented; the local-orbital
cutoff radius $R_{\phi}$ for each atomic species; and the density
kernel cutoff radius $R_{K}$.

Exchange and correlation is treated within the local density approximation
(LDA). Errors resulting from the supercell approximation, which can
be large in systems with a monopole or a strong dipole, are eliminated
using a truncated Coulomb potential.\cite{PhysRevB.73.205119,cutoff-coulomb}
Basis set superposition error that could affect the treatment, within
a local-orbital framework, of surface adsorption is eliminated by
the optimization procedure.\cite{bsse}

A further advantage of our method over other computational methods
that have been used to study nanocrystals,\cite{wang} is that the
whole of the nanostructure is included in the calculation in a way
which allows the electrons throughout the nanostructure to reach a
global equilibrium. We are therefore able accurately to account for
any coupling that may (and in fact does, as we shall show) occur between
different regions of the nanostructure. We caution that this method
presupposes integer occupations, which precludes partial occupancies
of states which might otherwise occur in a traditional calculation
where the system is treated as metallic. We have also performed test
calculations which permit fractional occupancies (albeit with cubic-scaling
computational cost) on representative smaller systems, which confirm
that the states presented here are indeed lowest in energy.

Primarily, we study nanorods of wurtzite GaAs (though we also model
GaN and AlN), since it exhibits all of the important characteristics
of a polar semiconductor i.e.\ elements of both ionic and covalent
bonding character and a non-centrosymmetric lattice structure. Ion
cores are represented using norm-conserving pseudopotentials. It has
been shown in previous work\cite{GaAs-nlcc} that an adequate description
of the geometry of systems containing Ga, requires either the explicit
inclusion of the Ga $3d$ electrons in the calculation, or, if the
$3d$ electrons are frozen into the pseudopotential, non-linear-core-corrections\cite{nlcc}
should be applied. To reduce the computational cost, we have chosen
the latter approach for both the Ga and As pseudopotentials.

An effectively infinite kernel cutoff radius $R_{K}$ was used in
order to treat insulators and metals on an equal footing. Calculations
using plane-wave DFT, as implemented in the \textsc{castep} code,\cite{castep}
show that setting $E_{\text{cut}}=400$~eV is sufficient to converge
bond-lengths, bond-angles and total energies of bulk GaAs, Ga$_{2}$
and As$_{2}$ dimers to within 0.02\% of their 800~eV values, using
our pseudopotentials. We find that bond-lengths are underestimated
by 1.3\%, which is typical for LDA. \textsc{onetep} is known to require
a $10$-$20\%$ larger $E_{\text{cut}}$ than \textsc{castep} for
the same level of convergence,\cite{onetep-forces} thus, the calculations
in this work use $E_{\text{cut}}=480$~eV and a generous local orbital
radius of $R_{\phi}=0.53$~nm.

For analysis of the dipole moment, we calculate the quantity $\mathbf{d}=-\int\! n(\mathbf{r})\mathbf{r}\, d\mathbf{r}+\sum_{I}Z_{I}\mathbf{R}_{I}$
from the density $n(\mathbf{r})$ in the whole simulation cell, and
the positions $\mathbf{R}_{I}$ of the ions of charge $Z_{I}$. The
internal electric field is calculated from the gradient of the value
of the local effective potential smoothed over a volume equivalent
to one primitive cell of the underlying material, as in our previous
work\cite{PhysRevB.83.241402}.

\section{Fermi level pinning in nanorods}

\label{sec:FLP}

\begin{figure}
\includegraphics[width=86mm]{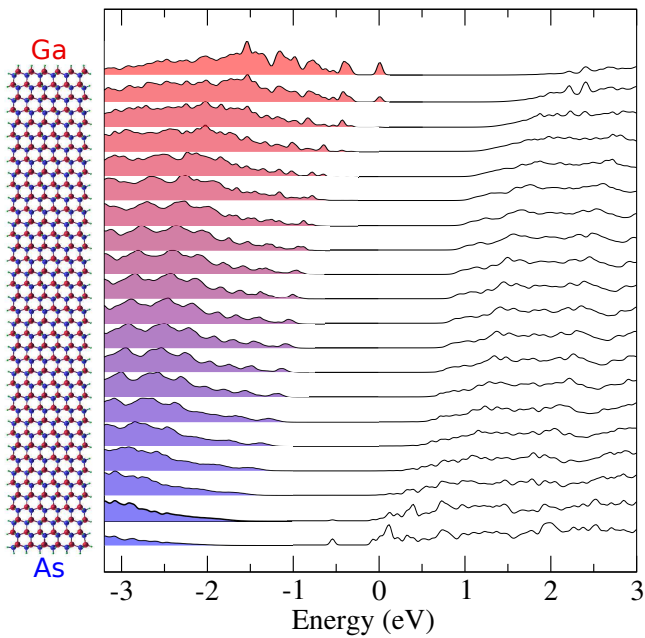} \caption{(Color online) Structurally relaxed, fully hydrogen terminated GaAs
nanorod (left) and the LDOS (right) for each `slab', consisting of
four planes of atoms (two As and two Ga). The filled curves indicate
the occupied (valence) states at each slab. The band-edge states at
opposite ends of the rod are seen to coincide in energy. }

\label{fig:H/H-r_LDOS} 
\end{figure}

We first consider the ground-state electronic structure of a structurally
relaxed nanorod of length 12.8~nm and cross-sectional area 3.56~nm$^{2}$,
comprising 2862 atoms. The rod (represented schematically in Fig.~\ref{fig:H/H-r_LDOS})
is labelled H/H-r, where the first three symbols (H/H) denote that
the lateral/end surfaces are terminated with hydrogen atoms, and `-r'
denotes that it is structurally relaxed.

This rod has a large negative dipole moment of $-600$~D and a large
internal field of $+0.1$~V/nm in the center of the rod. We adopt
the convention that a negative dipole moment is one whose direction
opposes that of the spontaneous polarization of the underlying wurtzite
crystal lattice (the wurtzite $[0001]$ direction, which is referred
to as the $z$ direction in this work). In Fig.~\ref{fig:H/H-r_LDOS}
we plot the `slab-wise' local density of electronic energy states
(LDOS) for this rod. We define a slab LDOS as follows: the rod is
nominally divided into 20 slabs along its length (the $z$-direction),
each consisting of four planes of atoms: two each of Ga and As. The
slab LDOS is the sum of the contributions to the total DOS from the
local orbitals centered on those atoms. In Fig.~\ref{fig:H/H-r_LDOS}
we superpose these slab LDOS. It is clear that the electric field
shifts the individual slab LDOS with respect to one another.

The Fermi energy can thus be considered to coincide with a significant
density of states on both polar surfaces of the nanorod. On the Ga(-H)
polar surface these states are mid-gap states, and on the As(-H) surface,
these mid-gap surface states are adjacent to the conduction band edge.
These are very stable positions for the Fermi level because small
deviations from these positions would cause changes in occupancy of
the surface states, resulting in a redistribution of charge and a
potential opposing the redistribution. This is analogous to Fermi
level pinning exhibited by some semiconductor surfaces, in which a
group of mid-gap states fixes the Fermi level at the surface at the
position of their average energy due to the action of surface states
as donors or acceptors, which get filled or emptied to compensate
for any change that may affect the relative position of the Fermi
level (e.g.\ the application of a voltage). We see this principle
in action in Fig.~\ref{fig:H/H-r_LDOS}, in that any significant
occupancy of the lowest-energy empty state on the As(-H) surface (which
appears to lie below the Fermi level) would in fact bring it above
the Fermi level due to the change in the electric field produced by
the charge redistribution. Of course, although this filling and emptying
of states can occur unaided in a DFT calculation, it would, in real
systems, depend on the availability of free charges in the environment,
implying an important role for the solvent.

There are at least two important differences between FLP on semi-infinite
surfaces and the finite end surfaces of nanorods; first, on surfaces
of area $A$, changes in surface charge density $\Delta\sigma$ due
to changes in occupancy of surface states come in discrete amounts
(i.e.\ $\Delta\sigma=e/A$), meaning that the continuous variability
of the surface charge density on semi-infinite surfaces gives way
to a discrete variability on finite surfaces; second, the analogue
of the \textit{depletion region} associated with FLP is the charged
region on the opposite end of the nanorod, meaning that the two surfaces
are coupled. This second effect may confer an important role on the
environment surrounding the nanorod, which may mediate the interaction
between the coupled ends by facilitating the transfer of electrons
between them as the system is perturbed.

In our previous work,\cite{PhysRevB.83.241402} we studied rods with
a range of different polar surface terminations, and with dipole moments
ranging from $+330$~D to -$614$~D. In all cases, the nanorods
exhibited this same feature of having Fermi levels coinciding with
the energies of large densities of mid-gap states on the end polar
surfaces of the nanorods. The arguments made here about FLP apply
to all nanorods with this feature.

One immediately obvious consequence of this picture is that the dipole
moment and internal field of a nanorod are dependent on the energies
of the pinning states on both ends of the rod, relative to their local
(slab) band edges. The difference between these relative energies
defines how much the energy spectrum is shifted between the top and
bottom ends of the nanorod i.e.\ the potential difference $\Delta V$
between the ends. If the Fermi level is pinned on both ends of the
rod, then the potential difference $\Delta V$ must also be pinned.
We will find, in each of the subsequent sections in this work, that
this pinning of $\Delta V$ plays a crucial role in determining the
polarity of a nanorod.

The pinning states in rod H/H-r on both ends of the rod are mid-gap
states, though they are adjacent to the band-edges in this case. Different
surface reconstructions on the polar surfaces may remove these mid-gap
states or change their positions relative to the local energy spectra.
This could change the potential difference across the rod and, therefore,
the dipole moment.

\section{Effect of surface chemistry on dipole moment}

\label{sec:ionic-charge}

Another implication of the picture presented above is that it is overly
simplistic to cast the problem of nanorod polarity in terms of an
ionic model, or a simple bond-electron counting model, since these
models do not include constraints on the potential difference across
a nanorod imposed by FLP. In previous work,\cite{PhysRevB.83.241402}
we found that the dipole moment $d_{z}$, the charge on the bottom
(As-rich) end $Q_{\text{b}}$, and the electric field in the middle
of the rod $E_{\text{m}}$ for two unrelaxed nanorods (labelled H/H
and H/P) were all very similar, despite having surface terminating
species of very different ionic charge. Rod H/H is fully hydrogen
terminated on both the lateral ($\parallel$ to $z$) surfaces and
the polar ($\perp$ to $z$) surfaces. Rod H/P, on the other hand,
is terminated with hydrogen atoms on the lateral surfaces, while on
the polar surfaces there are pseudo-hydrogen\cite{PhysRevB.71.165328}
atoms of two different varieties. These pseudo-hydrogen atoms are
used to passivate the dangling bonds of their respective surfaces:
those on the Ga polar surface have an ionic charge of $+1.25e$, while
those used to terminate the As polar surface have an ionic charge
of $+0.75e$. These pseudo-atoms are intended to passivate dangling
bonds on the polar surfaces, without adding charge to them, and they
have been shown in other work to render the surfaces electronically
inert.\cite{PhysRevB.71.165328}

A simple bond-electron and ion counting argument predicts that the
Ga polar surface on H/P should have an additional charge of $+0.25e$
for each of the 27 bound pseudo-atoms, compared to H/H -- a total
change of $+6.75e$ for each end. Similarly, the As polar surface
should have a reduced charge of $-0.25e$ per pseudo-atom -- a total
change of $-6.75e$. Nanorod H/P should therefore have a greatly reduced
dipole moment and potential difference across it. In fact, we observed
$d_{z}$, $Q_{\text{b}}$, and $E_{\text{m}}$ change from $-614$~D,
$1.00e$ and 0.100~V/nm respectively in H/H, to $-531$~D, $0.95e$
and 0.105~V/nm in H/P -- a much smaller change.

\begin{figure}
\includegraphics[clip,width=0.9\columnwidth]{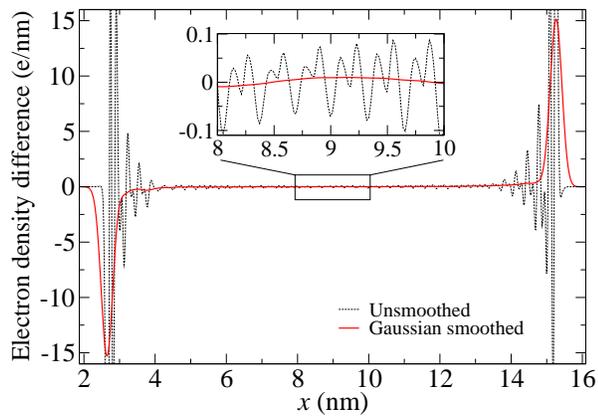} \caption{(Color online) The difference in laterally-integrated electron density
profile between H/H and H/P. The standard deviation of the Gaussian
used to smooth the data parallel to the nanorod axis is 0.32~nm.
There has been a shift of 6.70 electrons from left to right. We show
that the majority of this redistribution is attributable to changes
in surface state occupancy.}

\label{fig:HHminusHP} 
\end{figure}

\begin{figure}
\includegraphics[width=86mm]{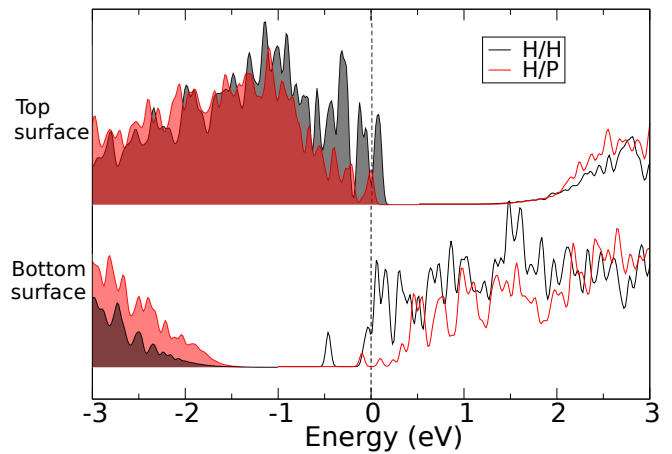} \caption{(Color online) Local densities of states for the slab of atoms on
the Ga-rich (top) and As-rich (bottom) ends of nanorods H/H and H/P.
The potential difference between the two ends $\Delta V\sim1.8$~eV
for H/H and $1.5$~eV for H/P.}

\label{fig:HH_HP_LDOS} 
\end{figure}

We plot the electron density difference between rods H/H and H/P in
Fig.~\ref{fig:HHminusHP}. The densities have been integrated in
the $x$- and $y$-directions and convolved with a Gaussian of standard
deviation 0.32~nm in the $z$-direction. The latter process smooths
out variations on length-scales smaller a unit cell length. By integrating
the resulting curve from each end to the center of the rod, we find
that there has been a transfer of 6.70 electrons from one end of the
rod to the other between rods H/H and H/P, which almost entirely cancels
the change in ionic charge. In Fig.~\ref{fig:HH_HP_LDOS}, we plot
the LDOS of only the top (the Ga rich polar end surface) and bottom
(the As rich polar end surface) slabs of both rods H/H and H/P. By
summing the occupations of the states plotted in this figure, we find
that there has been a change of six in the number of occupied states
on each end of the rod between H/H and H/P. The remaining charge transfer
of $0.70e$ must be associated with the polarization of occupied states
in slabs far away from the ends. This polarization of the electron
density can be observed in the inset to Fig.~\ref{fig:HHminusHP}.
The potential difference between the ends $\Delta V$ is very similar
for both rods -- $1.8$~eV for H/H and $1.5$~eV for H/P.

It is instructive to consider a fictitious adiabatic process in which
the ionic charge of the polar terminating species is slowly tuned
so as to go from rod H/P to H/H. The LDOS on the rod ends begins with
the Fermi level at the local band edge on each end of the rod, adjacent
to the electronic states. Therefore, we have $\Delta V\approx E_{\text{g}}$
in this case. As the charges of the terminating pseudo-atoms decrease
on the Ga end, and increase on the As end, the energy of nearby electronic
states on the Ga end of the rod must increase, pulling some of those
that lay just below the Fermi level, above it, and vice-versa on the
As end. This causes these states to change occupancy and compensate
some of the change in ionic charge. The higher the density of states
at the Fermi level, the less mobile is the Fermi level (i.e.\ the
more strongly the Fermi level is pinned). To effect a given shift
in the Fermi level, a larger change in surface ionic charge is required
if the density of states is high. That is to say, energies coinciding
with a high density of states (like the band edges) represent regions
of high stability for the Fermi level. The transition from H/P to
H/H causes the Fermi energy to run in to the (local) band edges, which
is why there is very little change in the pinned position of the Fermi
level on both ends, and therefore very little change in the potential
difference $\Delta V$ between the ends of the rod.

The general conclusion from this section is that changes in nanorod
polarity due to changes in ionic charge at the surfaces of nanorods
can be screened out due to FLP occurring at the ends of the nanorod.
This effect tends to preserve the potential difference between the
ends of the nanorod $\Delta V$, and consequently, the dipole moment.
The band-gap $E_{\text{g}}$, in effect, imposes an approximate upper
limit on $\Delta V$, since the density of states within the bands
is so high that the Fermi level would be very strongly pinned at its
edges.

\section{Effect of length and cross-sectional area on dipole moment}

\label{sec:NR-size}

In this section, we look at how the dipole moment of nanorod H/H varies
with rod length $L$, and cross-sectional area $A$, and show how
it can be explained using our FLP model.

\begin{figure}
\includegraphics[width=86mm]{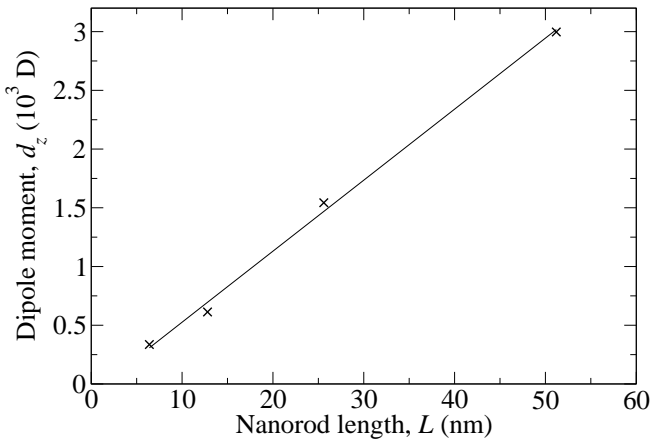} \caption{The magnitude of the dipole moment increases linearly with nanorod
length for nanorods of cross-sectional area $A=3.56\text{nm}^{2}$. }

\label{fig:dvslength} 
\end{figure}

\begin{figure}
\includegraphics[width=86mm]{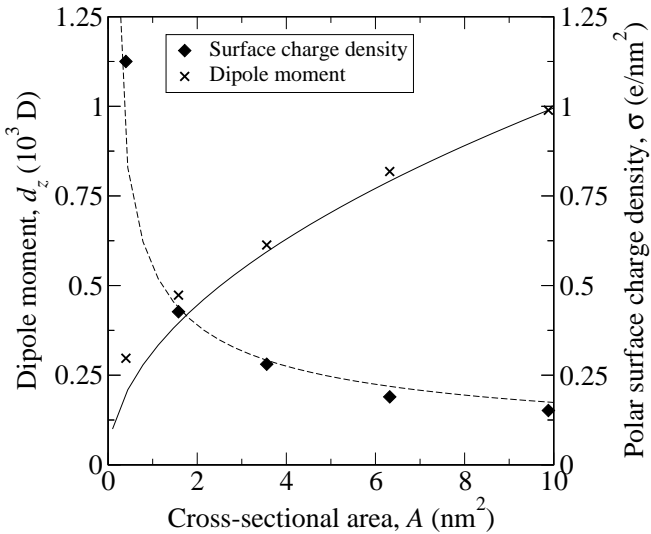}

\caption{The magnitude of the dipole moment increases with nanorod cross sectional
areas for nanorods of length $L=12.8\text{nm}$. Curves are fitted
to the data, with functional forms $\sigma(A)=c_{1}/(\sqrt{A}+c_{2}-\sqrt{A+c_{2}^{2}})$
and $d_{z}=c_{3}A/(\sqrt{A}+c_{2}-\sqrt{A+c_{2}^{2}})$, derived from
Eq.~\ref{eq:sigma}. Over this range $\sqrt{A}\ll c_{2}$ placing
these rods firmly in the {}``thin'' regime.}

\label{fig:RvsA} 
\end{figure}

We find, from Fig.~\ref{fig:dvslength}, that the dipole moment increases
roughly linearly with $L$ over the range studied, for rods of $A=3.56$~nm$^{2}$.
This implies that the excess polar surface ground-state charge density
on each end surface is independent of nanorod length over this range.

In Fig.~\ref{fig:RvsA} we show how both the dipole moment, and the
polar surface charge \textit{density} $\sigma$ on the bottom (As)
end surface of the rod, changes with $A$ for a fixed nanorod length
of $L=12.8$~nm. The charge density $\sigma$ on the polar end surfaces
decreases rapidly with cross-sectional area, asymptotically approaching
a constant value that may well be slightly above zero for nanorods
of this length (because surfaces of polar thin-films, unlike semi-infinite
surfaces, can support a non-zero charge\cite{Goniakowski}).

\begin{figure}
\includegraphics[width=86mm]{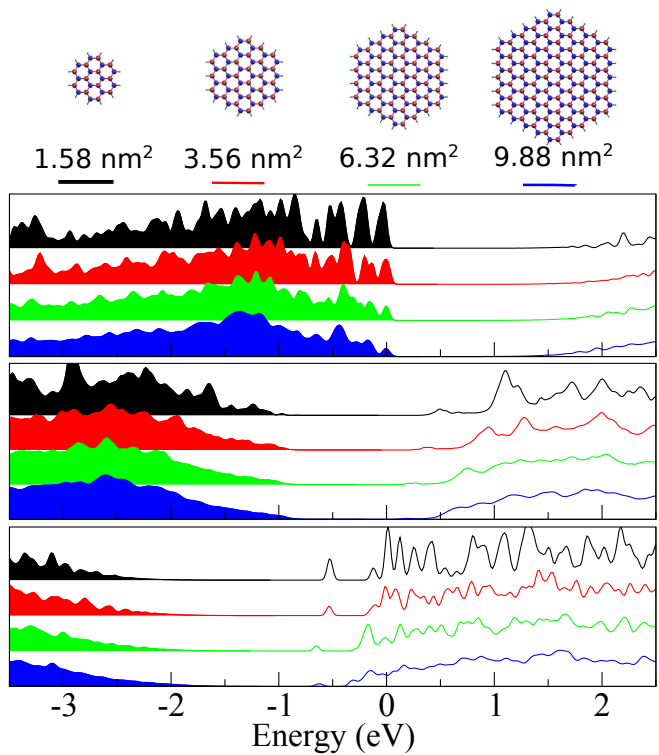} \caption{(Color online) Slab-wise local densities of states for rods of four
different cross-sectional areas, sampled at three positions on the
rod: (top) the slab on the Ga-rich end, (middle) the slab in the middle
of the rod, (bottom) the slab on the As-rich end. For ease of comparison,
we have shifted the energy of the highest occupied state for each
rod to zero. The Fermi level can be imagined to remain adjacent to
the band edges for all rods, and the band-gap is larger for thinner
rods due to quantum confinement.}

\label{fig:LDOSvsA} 
\end{figure}

We turn to consider the causes of these scaling relationships, focusing
first on the variation in rod polarization with respect to $A$. The
slab-LDOS plots in Fig.~\ref{fig:LDOSvsA} show that for all of the
cross-sectional areas studied, the occupied states on top surface
align closely with the unoccupied conduction band edge on the bottom
surface. In the previous section we argued that the local band-edges
represented an effective upper and lower limit for the Fermi energy
on the ends of a nanorod, and that the polarization of rod H/H, in
particular, is constrained by these band-edges (evidenced by the fact
that going from H/P to H/H does not change the dipole moment very
much, because the Fermi level touches the band-edges at both ends
of the rod).

In such a rod, the potential difference between its ends, $\Delta V$,
is determined mostly by its band-gap $E_{\text{g}}$, so that $\Delta V\approx E_{\text{g}}$.
We will argue that this observation alone can qualitatively account
for the observed trends in $d_{z}$ and $\sigma$ with $A$ in Fig.~\ref{fig:RvsA}.
While we do not expect the band-gap of equivalent real nanostructures
to match exactly with the DFT gaps we observe (due to the well-known
band-gap error of DFT), we expect qualitatively the same behavior
to emerge.

We can analyse this behaviour in terms of a simple electrostatic model,
and compare this to the results in Fig.~\ref{fig:RvsA}. The electrostatic
potential due to a circular disk of radius $a$ and area charge density
$\sigma$ at a distance $z$ along its axis is given by \begin{equation}
V(z)=2\pi\sigma\left(\sqrt{a^{2}+z^{2}}-|z|\right)\end{equation}
This expression simplifies to the familiar results for a point charge
in the limit that $z\gg a$ and infinite slab when $z\ll a$. Assuming
equal and opposite densities at the two ends of the rod, $z=0$ and
$z=L$, the total potential difference is $\Delta V=2\left[V(0)-V(L)\right]$
($\approx E_{\text{g}}$ in this case), which rearranges to give \begin{equation}
\sigma\approx\frac{E_{\text{g}}}{4\pi\left(a+L-\sqrt{a^{2}+L^{2}}\right)}\label{eq:sigma}\end{equation}
For {}``thick'' rods, $a\gg L$, $\sigma\sim E_{\text{g}}/L$, independent
of $a$ to leading order, whereas for {}``thin'' rods, $a\ll L$,
$\sigma\sim E_{\text{g}}/a\propto E_{\text{g}}/\sqrt{A}$. The rod
dipole moment $d_{z}=\sigma AL$ therefore scales as $d_{z}\sim E_{\text{g}}A$
for thick rods but as $d_{z}\sim E_{\text{g}}\sqrt{A}L$ for thin
rods. Substituting $a\propto\sqrt{A}$ into Eq.~\ref{eq:sigma} yields
a general expression for $\sigma(A)$, which we fit to the data in
Fig.~\ref{fig:RvsA}. We also fit the curve given by the expression
$d_{z}=\sigma AL$ to the data for $d_{z}$.

Evidence of deviation between our model and the data can be seen at
smaller values of $A$ in the data for $d_{z}$. The smaller $A$
is, the larger the error in our model. This is not surprising because
the model assumes that charge is localised on planes at the ends of
the rod, but we know that as $A$ becomes smaller, the surface charge
becomes increasingly delocalized along $z$. Furthermore, at small
$A$ the rod cross-section is increasingly dominated by edge atoms
rather than atoms truly belonging to the polar surface. For these
reasons, a breakdown of the model is expected at very small values
of $A$. Despite this complication it is clear from the fitting parameters
that our rods are in the {}``thin'' regime, as defined above, as
the model form correlates well with the observed behaviour. In summary,
thinner nanorods exhibit stronger decay of their internal potential,
due to finite width effects, therefore thinner rods require a larger
charge density on the nanorod ends in order to generate the required
potential difference $\Delta V$, than do thicker rods.

There is a second and less significant feature in the LDOS plots of
Fig.~\ref{fig:LDOSvsA}, that serves slightly to complicate the picture
described above. From the data sets in the middle window of Fig.~\ref{fig:LDOSvsA},
thinner nanorods are found to exhibit a larger local band-gap than
thicker rods. The local band-gap in the middle of the rod is found
to be 1.3~eV in the thinnest rod, and 0.9~eV in the thickest. This
is due to quantum confinement of electronic states in the lateral
direction, which is stronger in thinner rods. As the band-gap increases,
the potential difference between the ends of the rod can increase,
which further increases the amount of charge density required on the
end surfaces of the thinner nanorods in order to meet the resulting
increased pinned potential difference.

Although both of these effects (i.e.\ loss of the internal field
due to finite size effects, and the increase in the band-gap due to
quantum confinement) play a role in generating the behavior seen in
Fig.~\ref{fig:RvsA}, the first is more significant, since quantum
confinement produces only a 44\% increase in the band-gap over the
range of rods studied, which does not come close to accounting for
the 740\% increase in the polar surface charge density over the same
range.

We return now to the variation in nanorod polarization with $L$.
We did not observe quantum confinement related variation as was observed
over the range of $A$. Presumably, this is due to the large extent
of the rods in the $z$-direction. However, just like over the range
of $A$, we found that the Fermi level remains pinned close to the
band edges over the range of $L$, resulting in the potential difference
between the nanorod ends remaining constant.

The rods in Fig.~\ref{fig:dvslength} are able to maintain the charge
on their ends as $L$ increases, without incurring a significant change
in the potential difference across the rod because the rod is very
thin, and the internal potential decays very strongly: the field in
the middle of rods of length 12.8~nm, 25.6~nm and 51.2~nm are found
to decay to values of 0.1~V/nm, 0.035~V/nm and 0.009~V/nm respectively
in the rod centers. If the rods were thicker, we would expect this
decay to be weaker, and the amount of charge on the ends to be reduced
with $L$ to maintain the pinned potential difference, thus reducing
the rate at which the dipole moment increases with length.

In summary, FLP plays a determining role in the scaling of the dipole
moment of the nanorods studied, with length and cross-sectional area.
This effect manifests itself in different scaling behavior, the details
of which depend primarily on the rate of decay of the internal electric
field (which is a function of $A$), the length $L$ of the rods,
and the pinned potential difference $\Delta V$, which is close to
the size of the band-gap for rods in which the Fermi level is pinned
near the local band-edges, as is the case in the particular rods studied
in this section. Quantum confinement may also have some influence
on this scaling by affecting $\Delta V$.

\section{Effect of nanorod composition}

\label{sec:NR-type}

In this section we investigate how the polar behavior of nanorods
depends on composition. We calculate the charge distribution in three
rods -- one composed of GaAs, another of GaN, and a third of AlN.
These are all III-V semiconductors, so their chemistry and response
to terminating ligands can be expected to be similar. We therefore
terminate the rods with the same atoms as in previous sections, as
type H/P (lateral surfaces fully covered with hydrogen atoms, and
polar surfaces fully covered with the appropriate passivating pseudo-hydrogen
atoms). All have the same number of atoms (2862), and are constructed
of the same number of unit cells in each direction. Atoms are located
at their bulk equilibrium values, as calculated in the CASTEP plane-wave-DFT
code, meaning that the GaAs rod is longer than the GaN rod, which
in turn is longer than the AlN rod, because of the differences in
bulk lattice parameters.

The main characteristics of these rods and their charge distributions
are summarized in Table~\ref{tab:GaAs-GaN-AlN}, along with reference
information about the bulk properties of these semiconductors.

\begin{table}
 \begin{tabular}{rrrrr}
\hline 
 &  & AlN  & GaN  & GaAs\tabularnewline
\hline 
\multirow{6}{*}{Bulk} & DFT lattice param $a$ (\AA{})  & 3.075  & 3.154  & 3.935 \tabularnewline
 & DFT lattice param $c$ (\AA{})  & 4.941  & 5.132  & 6.486 \tabularnewline
 & DFT polarization (C/m$^{2}$)  & 0.073  & 0.029  & 0.005 \tabularnewline
 & DFT (LDA) bandgap (eV)  & 4.5  & 2.7  & 0.9 \tabularnewline
 & Experimental bandgap (eV)  & 6.2  & 3.3  & 1.5 \tabularnewline
 & Experimental permittivity, $\epsilon_{\text{r}}$  & 8.5  & 9.7  & 13.1 \tabularnewline
\hline 
\multirow{8}{*}{Rod} & Length, $L$ (nm)  & 9.66  & 10.01  & 12.61 \tabularnewline
 & Cross-sectional area, A (nm$^{2}$)  & 2.26  & 2.33  & 3.62 \tabularnewline
 & $d_{z}$~(D)  & -713  & -682  & -531 \tabularnewline
 & Polarization (C/m$^{2}$)  & -0.11  & -0.098  & -0.039 \tabularnewline
 & $\Delta V$~(eV)  & 4.2  & 3.2  & 1.5 \tabularnewline
 & $Q_{\text{b}}$~$(e)$  & 1.61  & 1.50  & 0.95 \tabularnewline
 & $\sigma_{\text{b}}$~$(e/\text{nm}^{2})$  & 0.711  & 0.645  & 0.262 \tabularnewline
 & $Q_{\text{b}}$ decay constant (nm$^{-1}$)  & 1.02  & 0.80  & 0.48 \tabularnewline
\hline
\end{tabular}\caption{Some properties of AlN, GaN, and GaAs in nanorod and in bulk. Experimental
data for AlN obtained from Refs.~\onlinecite{AlN-bg, AlN-bg2,AlN-perm},
for GaN obtained from Refs.~\onlinecite{GaN-bg,GaN-bg2,GaN-perm}
and for GaAs obtained from Ref.~\onlinecite{GaAs-bg, GaAs-perm}.}

\label{tab:GaAs-GaN-AlN} 
\end{table}

\begin{figure}
 \includegraphics[width=86mm]{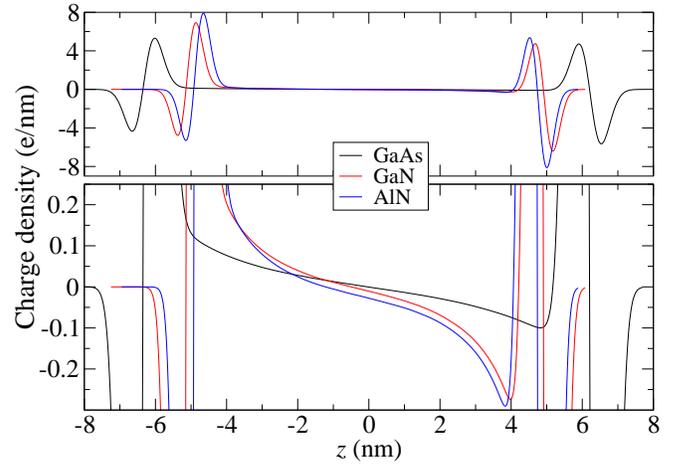}

\caption{(Color online) Laterally averaged and Gaussian-smoothed charge distributions
along the lengths of nanorods of AlN, GaN and GaAs. The ordinate has
been magnified in the lower panel.}

\label{fig:GaAs_vs_GaN_vs_AlN} 
\end{figure}

Figure~\ref{fig:GaAs_vs_GaN_vs_AlN} shows the distributions of charge
along the lengths of the rods for the three nanorods, integrated in
the $x$ and $y$ directions and in the $z$ direction, convolved
with a Gaussian of standard deviation $c/2$ so as to smooth out variations
on length-scales smaller than the length of half a unit cell length
$c$, (N.B. $c$ is different for each of the rods -- summarized in
Table~\ref{tab:GaAs-GaN-AlN}).

\begin{figure}
 \includegraphics[width=86mm]{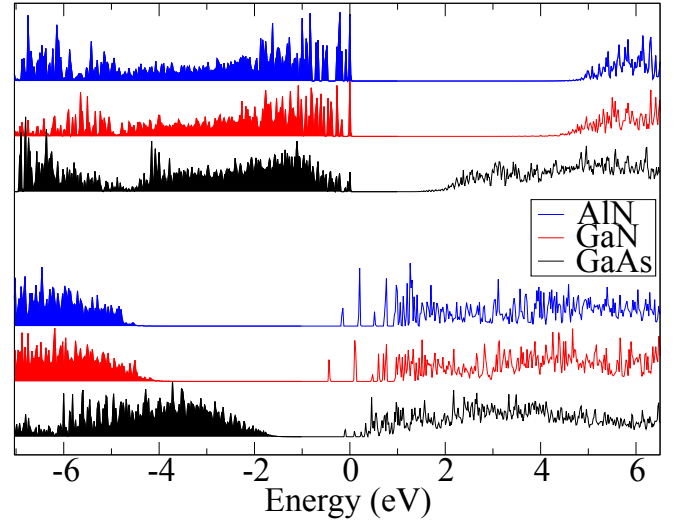} \caption{(Color online) Local densities of states of the cation-rich (top three
data sets) and anion rich (bottom three data sets) polar surfaces
of AlN, GaN, and GaAs. For ease of comparison, we have shifted the
energy of the highest occupied state of each rod to zero.}

\label{fig:GaN_AlN_LDOS} 
\end{figure}

In Fig.~\ref{fig:GaN_AlN_LDOS} we plot the LDOS for the polar surfaces
of the three rods. In all cases, the Fermi level can be imagined as
being pinned by surface states near the band-edges, for reasons outlined
in previous sections.

The polarization of the rod appears to increase proportionally with
the potential difference across the rod $\Delta V$, which is positively
correlated with the bulk semiconductor band-gap. The nanorod of the
largest band-gap semiconductor, AlN, supports the largest polarization,
and the nanorod with the lowest, GaAs, supports the smallest. However,
$\Delta V$ is not equal to, or proportional to, the bulk band-gap.
This is due to two factors: first, the effect of quantum confinement,
described in Sec.~\ref{sec:NR-size}, increases the band-gap by an
amount which varies depending on the type of material; and second,
the polar surface states responsible for pinning the Fermi level,
particularly on the bottom surface of the rod, can be seen in Fig.~\ref{fig:GaN_AlN_LDOS}
to lie at different positions relative to the local band edges in
all three rods.

The amount of excess charge on the bottom ends of the rods $Q_{\text{b}}$
is also positively correlated with the semiconductor band-gap. However,
it is not proportional to $d_{z}$, so there must be a significant
difference in how this charge is distributed along the rods. We measure
the decay rate of the long-range tails of excess charge which can
be seen in the magnified plot in Fig.~\ref{fig:GaAs_vs_GaN_vs_AlN}.
Nanorods of higher band-gap materials exhibit a larger decay constant
(Table~\ref{tab:GaAs-GaN-AlN}), and therefore, stronger localization
of their excess surface charges. This stronger localization is indicative
of the fact that rods of lower permittivity materials more strongly
concentrate the field lines associated with surface charge, and therefore
exhibit a weaker long-range decay of their internal electric fields
for a given finite cross-sectional area. Therefore, rods of lower
permittivity materials require less excess charge density on their
ends to attain a particular potential difference $\Delta V$ (and
polarisation), than do rods of higher permittivity materials. This
is a similar argument to the one in Sec.~\ref{sec:NR-size}, which
also concluded that rods exhibiting weaker decay of their internal
fields (i.e.\ thick rods), require less excess surface charge density
to attain a particular $\Delta V$. This effect can be partially incorporated
in to our model in Sec.~\ref{sec:NR-size}, by introducing a material-dependent
constant of proportionality which determines the effective cross-sectional
area seen by the electrons, for a given geometrical cross-sectional
area. This effective cross-sectional area is larger in materials of
lower permittivity.

\section{Summary and conclusion}

\label{sec:conclusion}

The potential difference across a nanorod due to its large dipole
moment shows up in the LDOS as a shifting of the energy of the states
as one moves along the length of the rod. In this work and in our
previous work,\cite{PhysRevB.83.241402} it has been found that nanorods
of a variety of surface terminations have Fermi levels which coincide
with a high LDOS at their polar end surfaces. These are either mid-gap
states or states close to the band-edges. In the latter case, this
means that the potential difference across the rod is approximately
equal to its local band-gap.

These are very stable positions for the Fermi level because small
deviations from these energies result in changes in occupancy and
a redistribution of charge, which generates a potential that opposes
the initial change. This phenomenon is a generalization of the FLP
effect on semi-infinite surfaces to structures of small dimensions.

In this work, we provide evidence that FLP plays a determining role
for the polarity of nanorods. Pinning of the Fermi level results in
a pinning of the potential difference $\Delta V$ across the nanorod,
and hence its dipole moment.

We demonstrate that simple ionic or bond-electron counting models
can be inadequate for describing, even qualitatively, differences
in polarity between nanorods of different surface termination. In
particular, we have shown that the effect of varying the ionic charge
on the ends of a rod can be screened out, due to pinning at the nanorod
ends, so as to maintain its polarity.

We show that FLP can play a determining role for the scaling of the
dipole moment with nanorod size. It is also able to account for differences
in polarity between nanorods of different composition.

A particularly striking consequence of this effect is that it implies
a crucial role for the solvent in determining the properties of a
nanorod. Not only does the choice of solvent determine whether charge
can be transferred between the ends of the nanorod, because it mediates
this transfer, but it can also alter the LDOS on the nanorod ends
by changing the surface chemistry. We propose that this latter effect,
coupled with FLP, could have a dramatic effect on the dipole moment,
and hence the optical properties.

Clearly, the picture discussed in this work could have important consequences
for the response properties of nanorods in applied electric fields,
and in the fields of neighboring polar nanorods. This could be important,
not only for their optical properties, but also for the energetics
of self-assembly of polar semiconductor nanostructures.

This work was supported by EPSRC (UK) under Grant No. EP/G05567X/1,
the EC under Contract No. MIRG-CT-2007- 208858, and a Royal Society
University Research Fellowship (PDH). All calculations were run on
the Imperial College HPC Service.


\begin{thebibliography}{36}
\bibitem{smallisdifferent}M.~A.~El-Sayed, Accounts Chem. Res. \textbf{37},
326 (2004). 

\bibitem{X.Michalet01282005}X.~Michalet, F.~F.~Pinaud, L.~A.~Bentolila,
J.~M.~Tsay, S.~Doose, J.~J.~Li, G.~Sundaresan, A.~M.~Wu, S.~S.~Gambhir,
and S.~Weiss, Science \textbf{307}, 538 (2005). 

\bibitem{NirTessler02222002}N.~Tessler, V.~Medvedev, M.~Kazes,
S.~Kan, and U.~Banin, Science \textbf{295}, 1506 (2002). 

\bibitem{kazesetal}M.~Kazes, D.~Y.~Lewis, Y.~Ebenstein, T.~Mokari,
and U.~Banin, Adv. Mater. \textbf{14}, 317 (2002). 

\bibitem{Wendy}W.~U.~Huynh, J.~J.~Dittmer and A.~P.~Alivisatos,
Science \textbf{295}, 2425 (2002). 

\bibitem{Nieetal}Z.~Nie, A.~Petukhova, and E.~Kumacheva, Nat.
Nanotechnol. \textbf{5}, 15 (2010). 

\bibitem{shevchenko} E.~V.~Shevchenko, D.~V.~Talapin, N.~A.~Kotov,
S.~O'Brien, and C.~B.~Murray, Nature \textbf{439}, 55 (2006). 

\bibitem{PhysRevLett.79.865}S.~A.~Blanton, R.~L.~Leheny, M.~A.~Hines,
and P.~Guyot-Sionnest, Phys. Rev. Lett., \textbf{79}, 865 (1997); 

\bibitem{shim:6955}M.~Shim and P.~Guyot-Sionnest, J. Chem. Phys.
\textbf{111}, 6955 (1999). 

\bibitem{PhysRevLett.90.097402}L.-S.~Li and A.~P.~Alivisatos,
Phys. Rev. Lett. \textbf{90}, 097402 (2003); 

\bibitem{talapin} D.~V.~Talapin, E.~V.~Shevchenko, C.~B.~Murray,
A.~V.~Titov, and P.~Král, Nano Lett. \textbf{7}, 1213 (2007). 

\bibitem{Goniakowski} J.~Goniakowski, F.~Finocchi, and C.~Noguera,
Rep. Prog. Phys. 71, 016501 (2008) 

\bibitem{rabani:5355}E.~Rabani, B.~Hetényi, B.~J.~Berne, and
L.~E.~Brus, J. Chem. Phys. \textbf{110}, 5355 (1999). 

\bibitem{rabani:1493}E.~Rabani, J. Chem. Phys. \textbf{115}, 1493
(2001). 

\bibitem{Shanbhag:2006ix}S.~Shanbhag, and N.~A.~Kotov, J. Phys.
Chem. B \textbf{110}, 12211 (2006). 

\bibitem{wang}S.~Dag, S.~Wang, and L.-W.~Wang, Nano Lett. \textbf{11},
2348 (2011). 

\bibitem{PhysRevB.83.241402}P.~W.~Avraam, N.~D.~M.~Hine, P.~Tangney,
and P.~D.~Haynes, Phys. Rev. B \textbf{83}(\textbf{24}), 241402
(2011). 

\bibitem{skylaris:084119}C.-K.~Skylaris, P.~D.~Haynes, A.~A.~Mostofi,
and M.~C.~Payne, J. Chem. Phys. \textbf{122}, 084119 (2005). 

\bibitem{onetep-forces}N.~D.~M.~Hine, M.~Robinson, P.~D.~Haynes,
C.-K.~Skylaris, M.~C.~Payne, and A.~A.~Mostofi, Phys. Rev. B
\textbf{83}, 195102 (2011). 

\bibitem{onetep-pwaccuracy}C.-K.~Skylaris and P.~D.~Haynes, J.
Chem. Phys. \textbf{127}, 164712 (2007). 

\bibitem{PhysRevB.73.205119}C.~A.~Rozzi, D.~Varsano, A.~Marini,
E.~K.~U.~Gross, and A.~Rubio, Phys. Rev. B \textbf{73}, 205119
(2006). 

\bibitem{cutoff-coulomb}N.~D.~M.~Hine, J.~Dziedzic, P.~D.~Haynes,
and C.-K.~Skylaris, J. Chem. Phys. \textbf{135}, 204103 (2011).

\bibitem{bsse} P.~D.~Haynes, C.-K.~Skylaris, A.~A.~Mostofi and
M.~C.~Payne, Chem. Phys. Lett. \textbf{422}, 345 (2006). 

\bibitem{GaAs-nlcc} A.~Qteish and R.~J.~Needs, Phys. Rev. B \textbf{43}
4229 (1991) 

\bibitem{nlcc} S.~G.~Louie, S.~Froyen, and M.~L.~Cohen, Phys.
Rev. B \textbf{26} 1738 (1982) 

\bibitem{castep} S.~J.~Clark, M.~D.~Segall, C.~J.~Pickard,
P.~J.~Hasnip, M.~I.~J.~Probert, K.~Refson, and M.~C.~Payne,
Z. Kristallogr. \textbf{220} (2005) 567570 

\bibitem{0953-8984-17-37-012}C.-K.~Skylaris, P.~D.~Haynes, A.~A.~Mostofi,
and M.~C.~Payne, J. Phys.: Condens. Matter \textbf{17}, 5757 (2005). 

\bibitem{PhysRevB.71.165328}X.~Huang, E.~Lindgren, and J.~R.~Chelikowsky,
Phys. Rev. B \textbf{71}, 165328 (2005). 

\bibitem{AlN-bg} H.~Yamashita, K.~Fukui, S.~Misawa, and S.~Yoshida,
J. Appl. Phys. \textbf{50} (1979) 896. 

\bibitem{AlN-bg2} D.~Brunner, H.~Angerer, E.~Bustarret, F.~Freudenberg,
R.~Hopler, R.~Dimitrov, O.~Ambacher, and M.~Stutzmann, J. Appl.
Phys. \textbf{82}, 5090 (1997) 

\bibitem{AlN-perm} Y.~Goldberg, \textit{Properties of Advanced Semiconductor
Materials GaN, AlN, InN, BN, SiC, SiGe } . Eds. M.~E.~Levinshtein,
S.~L.~Rumyantsev, M.~S.~Shur, John Wiley \& Sons, Inc., New York,
2001, 31-47 

\bibitem{GaN-bg} H.~Teisseyre, P.~Perlin, T.~Suski, I.~Grzegory,
S.~Porowski, J.~Jun, A.~Pietraszko, and T.~D.~Moustakas, J. Appl.
Phys. \textbf{76}, 2429 (1994) 

\bibitem{GaN-bg2} I.~Vurgaftman and J.~R.~Meyer, J. Appl. Phys.
\textbf{94}, 3675 (2003). 

\bibitem{GaN-perm} V.~Bougrov, M.~E.~Levinshtein, S.~L.~Rumyantsev,
and A.~Zubrilov, \textit{Properties of Advanced Semiconductor Materials
GaN, AlN, InN, BN, SiC, SiGe.} Eds. M.~E.~Levinshtein, S.~L.~Rumyantsev,
M.~S.~Shur, John Wiley \& Sons, Inc., New York, 2001, 1-30. 

\bibitem{GaAs-bg} H.~C.~Casey, D.~D.~Sell, and K.~W.~Wecht,
J. Appl. Phys. \textbf{46}, 250 (1975) 

\bibitem{GaAs-perm} S.~Adachi, J. Appl. Phys, \textbf{53}, 12, 8775-8792,
1982
\end{thebibliography}
\end{document}